\documentstyle[10pt,psfig,aaspp4]{article}

\newcommand{\be}{\begin{equation}}
\newcommand{\ba}{\begin{eqnarray}}
\newcommand{\ee}{\end{equation}}
\newcommand{\ea}{\end{eqnarray}}  
\newcommand{\etal}{et al.\ }
\received{}   
\accepted{}   
\journalid{}{}
\articleid{}{}

\lefthead{Thacker, Scannapieco \& Davis}
\righthead{Violence in the Dark Ages}

\begin{document}

\def\lesssim{\mathrel{\hbox{\rlap{\hbox{\lower4pt\hbox{$\sim$}}}\hbox{$<$}}}}
\def\gtrsim{\mathrel{\hbox{\rlap{\hbox{\lower4pt\hbox{$\sim$}}}\hbox{$>$}}}}
\def\ApJ{{\em Astrophys.\ J.\ }}
\def\AJ{{\em Astron.\ J.\ }}
\def\MNRAS{{\em Mon.\ Not.\ R.\ Astron.\ Soc.\ }}
\def\CIV{\hbox{C$\scriptstyle\rm IV\ $}}\

\title{Violence in the Dark Ages}
\author{Robert J. Thacker\altaffilmark{1,2}, Evan 
Scannapieco\altaffilmark{3,2}}
\authoremail{thacker@physun.physics.mcmaster.ca}
\and
\author{Marc Davis\altaffilmark{2}}

\affil{${}^{1}$Department of Physics and Astronomy,
McMaster University, 1280 Main St. West, Hamilton, Ontario, L8S 4M1,
Canada.}
\affil{${}^{2}$Department of Astronomy,
University of California at Berkeley,
Berkeley, CA, 94720.}
\affil{${}^{3}$Osservatorio Astrofisico di Arcetri, Largo E. Fermi 5,
Firenze, Italy.}

\begin{abstract}
A wide range of observational and theoretical arguments suggest that the
universe experienced a period of heating and metal enrichment, most likely from
starbursting dwarf galaxies. Using a hydrodynamic simulation we have conducted
a detailed theoretical investigation of this epoch at the end of the
cosmological ``dark ages''. Outflows strip baryons from pre-viralized halos
with total masses $\lesssim$10${}^{10}$ M${}_\odot$, reducing their number
density and the overall star formation rate, while pushing these quantities
toward their observed values. We show that the metallicity of
$\lesssim$10${}^{10}$ M${}_\odot$ objects increases with size, but with a large
scatter, reproducing the metallicity-luminosity relation of dwarf galaxies.
Galaxies $\gtrsim$10${}^{10}$ M${}_\odot$ form with a roughly constant initial
metallicity of 10\% solar, explaining the observed lack of metal-poor disk
stars in these objects. Outflows enrich roughly 20\% of the simulation volume,
yielding a mean metallicity of 0.3\% solar, in agreement with observations of
\CIV in QSO absorption-line systems.
\end{abstract}

\section{Introduction}
An epoch of galaxy outflows is a virtually inevitable consequence of
hierarchical structure formation.  In such models, subunits merge and
accrete mass to form objects of ever-increasing size with a rich
substructure.  Yet while the sizes of galaxies increase strongly over
time, the sizes of the structures within them remain largely fixed.
Thus such properties as the scale of associations of O and B 
stars (\cite{mc97}) and the energy deposition due to the resulting 
supernovae
(SNe) are likely to be weak functions of the mass of the host galaxy,
depending mainly on environmental effects such as the distribution of
stellar material, gas, and dust (\cite{om00}).  Hence, while the ejecta
of conglomerations of Type-II SNe are primarily confined to the
interstellar medium in Milky Way size galaxies, they are easily able
to escape from the dwarf galaxies whose formation at high-redshift
ushered in the first period of intense star formation in the universe.
Thus the end of the dark ages, before galaxies formed, is likely to
have been accompanied by a potentially drastic change of both the
properties of the Intergalactic Medium (IGM) and the evolution of
collapsing protogalactic dwarf systems (Scannapieco, Thacker \& Davis 2001). 

From an observational point of view, the evidence for such an epoch is
equally compelling.  Analyses of \CIV lines in QSO absorption spectra
uncover an IGM that has been widely and inhomogeneously enriched by stellar
material (\cite{co91}), and heating from outflows is necessary to avoid
overproduction of the X-ray background (\cite{pe99}). Such mechanisms are
believed to be needed to explain the X-ray luminosity-temperature relation 
($L_x:T$)
of the hot gas in galaxy clusters, which is widely discrepant with
predictions that consider only heating from gravitational collapse
(\cite{ka91}).  Note that a recent argument by Voit \& Bryan (2001) shows
that the precise details of outflow heating are comparatively unimportant to 
the
shape of the $L_x:T$ relation (excessively heated gas will float to the 
edges of
a cluster where it is undetectable, further, simulations by Pearce \etal
(2000) demonstrate that cooling alone will produce a break in the $L_x:T$
relation). Finally, the starbursting galaxies that form the low-redshift end
of this epoch have been detected directly, both in optical and infrared
observations (\cite{pe01}) at $3 \lesssim z \lesssim 4$, and optical
observations (\cite{fb01}; \cite{hu02}) of lensed galaxies at $4 \lesssim z
\lesssim 6.56$.

Motivated by this wide range of observational evidence and theoretical
pointers, we have carried out a detailed numerical study of galaxy
outflows, their impact on the IGM, and the formation of galaxies within
it.  While previously studies of outflows and cosmological metal
enrichment have been attempted, these have either employed approximate
semi-analytical methods (\cite{nc95}; Madau, Ferrara \& Rees 2001; \cite{sc01}),
low-resolution simulations (\cite{a01a}; \cite{a01b}; \cite{co99}) that
are unable to identify the dwarf galaxies that are likely to be the
primary source of metals (\cite{mad01}; \cite{sc01}; Scannapieco, Ferrara \& 
Madau 2002), or a
high-resolution scheme in which supernovae explosions were not adequately
accounted for (\cite{go97}). We also conducted an earlier study
(\cite{st01}) in which we demonstrated that outflows can reduce the number
density of dwarf galaxies at high redshift. This effect was seen to be a
combination of `baryonic stripping', whereby outflows strip gas from low
over-density halos, and a delay in formation times caused by outflows
slowing accretion on to the host halo. However, due to the significant 
wall clock time required to conduct the simulations we were unable to 
integrate them beyond a redshift of $z=6$. In this paper we 
build 
on our previous work, firstly by improving the physical model through the
inclusion of variable metallicity and, secondly, we incorporate a new
numerical technique to integrate to a directly observable epoch.
By adopting a simple model for outflow generation and enrichment with
stellar material in high-resolution cosmological simulations, we
demonstrate that the of inclusion winds is able to explain a diverse set
of poorly understood properties of objects ranging from moderately-sized
galaxies to the absorption clouds that trace the most tenuous
cosmological structures.

\section{Simulations of Galaxy Outflows}

Based on the latest measurements of the Cosmic Microwave Background
and the number abundance of galaxy clusters (\cite{ba00}; \cite{vi96}) we
focus our attention on a Cold Dark Matter cosmological model with
parameters $h=0.65$, $\Omega_0$ = 0.35, $\Omega_\Lambda$ = 0.65,
$\Omega_b = 0.05$, $\sigma_8 = 0.87$, and $n=1$, where $h$ is the
Hubble constant in units of 100 km s$^{-1}$ Mpc$^{-1}$, $\Omega_0$,
$\Omega_\Lambda$, and $\Omega_b$ are the total matter, vacuum, and
baryonic densities in units of the critical density, $\sigma_8^2$ is
the variance of linear fluctuations on the $8 h^{-1}{\rm Mpc}$ scale,
and $n$ is the ``tilt'' of the primordial power spectrum.

As in our earlier work (\cite{st01}), simulations were conducted with a
parallel OpenMP based implementation of the ``HYDRA'' code (\cite{th00})  
that uses the Adaptive Particle-Particle, Particle-Mesh algorithm
(\cite{cu91})  to calculate gravitational forces, and the Smoothed Particle
Hydrodynamic (SPH) method (\cite{lu77}; \cite{gm77}) to calculate gas
forces.  We simulate a periodic box of size 5.2 $h^{-1}$ comoving Mpc
containing $192^3$ dark matter and $192^3$ gas particles with masses of $2.5
\times 10^6 M_\odot$ and $5.0 \times 10^5 M_\odot$ respectively. This is
sufficient to resolve the smallest objects that are able to cool efficiently
without molecular hydrogen, with masses $\gtrsim 10^8 M_\odot$.  We chose
this limit as primordial molecular hydrogen is extremely fragile, and is
likely to be quickly dissociated by the first stars that form at very high
redshift (\cite{ha97}).  Note that while alternative schemes for creating
cosmological H${}_2$ have been suggested, star formation and outflow
generation is likely to be inefficient in objects that rely on molecular
cooling, causing them to have a small overall impact on IGM heating and
enrichment (\cite{mad01}; \cite{ef02}).  Forces are softened using a fixed
physical Plummer softening length of 1.54 kpc, yielding a minimum
hydrodynamic resolution of 1.8 kpc; gas densities and energies are
calculated using the standard SPH smoothing kernel method (\cite{st01}),
with the kernel tuned to smooth over 52 particles; and radiative cooling is
calculated using standard tables (\cite{su93}).  Because the epoch of
reionization is unknown, and also for reasons of simplicity, we do not
include a fiducial photoionization background in the simulation.

Under heavy particle clustering, and when implemented with a minimum
resolution limit, the SPH method can degenerate to an order $N^2$
algorithm. While some authors resort to merging particles to avoid this
problem, this breaks the one-to-one relationship between mass
and particle number and leads to a degradation of accuracy in the
hydrodynamic calculation. Our solution to this problem (\cite{th01}) is to
calculate an aggregate solution in regions below the resolution limit,
grouping particles in sets of 4-30 that may change from step to step,
while never actually merging particles together. The errors introduced by
this method are far smaller than those created by merging particles, and
the algorithm remains less than order $N^{3/2}$ under heavy clustering,
which translates to a three-fold performance improvement over our previous
code (\cite{st01}).

To study the propagation of galaxy outflows, it is first necessary to
identify where galaxies are forming.  Typically, this is done using
the friends-of-friends group finding algorithm which relies upon
calculating inter-neighbor distances (\cite{da85}), however this
technique is very computationally expensive.  We therefore adopt an
alternative approach in which we search for regions where gas
particles exceed a density threshold $\delta_c$ times the mean gas
density.  Provided more than $2.6\times 10^7 M_\odot$ (our effective
mass resolution limit) is contained in such a region, the object is
counted as a galaxy.  This method performs well in comparison to the
friends-of-friends technique (\cite{st01}), and throughout this paper we
set $\delta_c =500.$

Star formation is modeled by converting a fixed fraction,
$\epsilon_{\rm sf}=0.1$, of the gas in a collapsed object into stars
in a single starburst.  Subsequent starbursts within a galaxy are
likely to be smaller and less efficient, so our model is a
conservative lower limit on star formation activity. The gas involved
in a starburst is tagged to prevent further bursts until a
sufficiently large reservoir of `fresh' gas is accumulated. Star
formation can only occur when over one-third of the galaxy, and at
least 52 new particles, are untagged, although testing has shown that
these thresholds can be varied over a wide range with only a small
effect on our results (\cite{st01}).  Finally, we also impose an upper
limit of 1000 particles per burst, corresponding to $5 \times 10^8
M_\odot$ of stars, which prevents unphysically large groups of
supernovae from exploding simultaneously and generating enormous
outflows (\cite{fp01}).

The structure of galaxy outflows has been studied in detail 
previously (\cite{ml99}; \cite{ma00}; Mori, Ferrara \& Madau 2001), yielding the 
conclusion that in
starbursting disk galaxies with masses $\lesssim 10^{10} M_\odot$
``blow-out'' occurs.  In these galaxies, the superbubbles around
groups of Type-II SNe punch out of the plane of the disk and shock the
surrounding IGM although they fail to excavate the interstellar medium
itself. Guided by this scenario, we model the sub-resolution physics
of outflow generation by rearranging the IGM surrounding a galaxy
while leaving the central object intact.

Hence we construct our outflows from gas particles within a radius
twice that of the central galaxy ($r \leq 2 r_g$), but outside the
galaxy itself ($r > r_g$). This local IGM is arranged randomly into
two concentric spherical shells of radius $2.0 r_g$ and $1.8 r_g$ of
equal mass.  This multi-shell structure assures that the outflows will
be sufficiently well-resolved radially to be reasonably treated by the
SPH algorithm. Sufficient resolution in each of the shells is assured
by arranging the particles in an anti-correlated fashion such that all
particles have an inter-neighbor distance in excess of half the
average inter-particle spacing.  We have tested this approach in
detail against one-dimensional analytical models of expanding shells,
and have found that it reproduces these results almost exactly 
(\cite{st01}).

The particles in these shells are assigned velocities that conserve
their overall center of mass velocity and angular momentum, and
include a velocity boost in the radial direction.  The strength of
this boost is determined by assuming that a fixed fraction of the
mechanical energy from supernovae from each starburst is channeled
into the outflow, $\epsilon_{\rm wind}=0.1$.  We estimate this kinetic
input to be $2\times 10^{51}$ ergs per supernova, to take into account
the contribution from stellar winds, and assume that one supernova
occurs for every 100 $M_\odot$ of stars formed (\cite{gi97}).

Our previous simulations side-stepped the issue of metal enrichment,
taking a fixed value of $0.05 \; Z_\odot$ for all gas particles.  In
this work we adopt a more realistic model in which the primordial
metallicity is taken to be zero, and gas particles are enriched by
ejecta from Type-II SNe. The metal content of each gas particle is
therefore allowed to evolve in a similar fashion to hydrodynamic
variables. We assume a standard yield of 2 $M_\odot$ of metals per
SNe, corresponding to an effective metal yield $p=0.02$, and
distribute this material equally between the galaxy and the outflow.
For these parameters a galaxy will be enriched following the first
outflow event to a metallicity $Z_{gal}$ given by,
$\log_{10}(Z_{gal}/Z_\odot) \sim \log_{10}(\epsilon_{\rm sf}\times f
\times p/Z_\odot)=-1.3$, if we take $f=0.5$ as the fraction of metals
retained by the galaxy.  Following an outflow event, the total
metal yield, by mass, for the galaxy is distributed evenly among 
gas
particles that determined to be in the galaxy ($\delta 
M_{particle}^Z=(1/N_{gal})f\,M_{total}^Z$, where $M_{total}^Z$ is the 
total metal yield, by mass, of the starburst, $N_{gal}$ is the number of 
particles 
in the galaxy and $\delta
M_{particle}^Z$ is the increase in the metal mass of a particle in 
the galaxy). Similarly, 
the metal yield deposited in the outflow is distributed evenly among gas
particles determined to be in the outflow ($\delta
M_{particle}^Z=(1/N_{outflow})(1-f)\,M_{total}^Z$). Individual gas 
particles can
undergo multiple episodes of enrichment in response to a series of
outflows.

Enriched particles retain their metals for the remainder of the
simulation, and the spatial metal distribution is calculated by
convolving with the SPH smoothing kernel. The smoothed field is then
used self-consistently in computing gas cooling. This may be viewed as
assuming the metals within a given particle's smoothing length are
well mixed, a reasonable approximation given that the Rayleigh-Taylor
instability will mix gases in a time roughly proportional to the size
of the region divided by the sound speed.

\section{Outflows and Structure Formation}

We have conducted two simulations: a fiducial run
in which we fix the star-formation and outflow efficiencies to be
$\epsilon_{\rm sf} = \epsilon_{\rm wind} = 0.1$, and a comparison
``no-outflows run'' in which we fix $\epsilon_{\rm sf} = 0.1$ but do
not include metal enrichment or outflows.  In the fiducial run the gas
is initially free of metals, while in the no-outflows run the
metallicity is fixed at $0.05 \;Z_\odot$ at all times.  While
particle-particle interactions halted our previous simulations at $z =
6$, the pseudo-merging scheme discussed above allowed us to reach $z=4$
in this study.  A total of 4515 outflows were generated in this case,
with initial radial velocities ranging from 80 to 300 km/s,
consistent with direct observations (\cite{pe01}; \cite{fb01}).

\begin{figure}[t]
\vspace{10cm}
\includegraphics{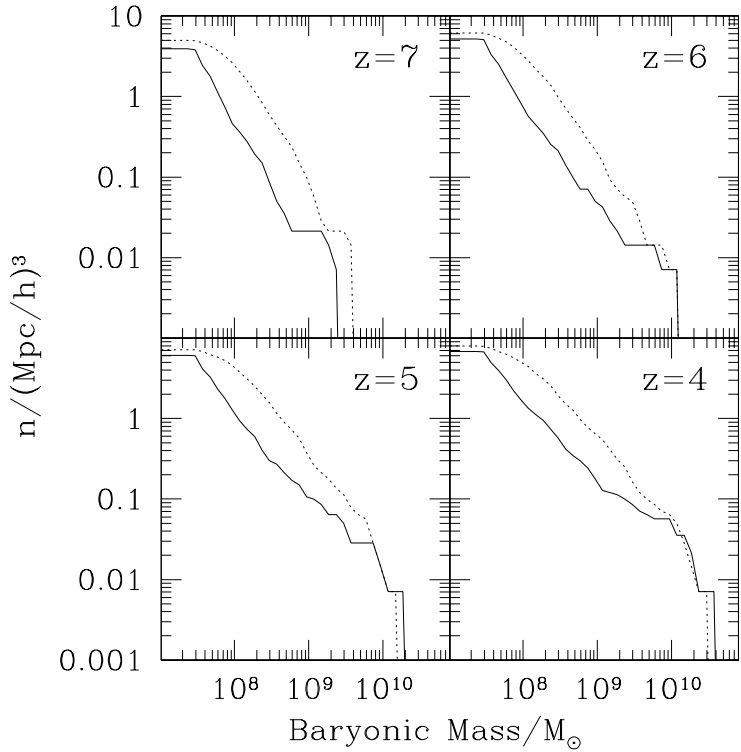}
\caption{Number density of galaxies as a function of their baryonic
mass. Redshift is given in the upper right corner of each panel, the
solid lines are from the simulation with
outflows, and the dotted lines are from the star-formation only 
simulation.}
\label{fig:fig1}
\end{figure}

The cumulative number density of galaxies within the simulations as a
function of baryonic mass and redshift is shown in Figure
\ref{fig:fig1}. Below a limiting mass of $5\times 10^9 M_\odot$, two
significant features are observed.  Firstly, there is an overall
reduction in the baryonic mass of the objects in the outflows
run. This is caused by outflows suppressing accretion onto the host
galaxy, a process that is similar to that described in earlier 
work (\cite{de86}) on dwarf galaxies
which showed that supernova-driven winds result in a systematic bias
toward higher mass-to-light ratios. Secondly, the
fiducial run also contains many examples of the ``baryonic stripping''
mechanism previously identified (\cite{sc00}).  In these cases, the
momentum in an outflow is sufficient to evacuate gas from nearby
overdense regions that would have otherwise later formed into dwarf
galaxies, leaving behind empty `dark halos' of cold dark matter.
Note also that this overall suppression of objects points to a
possible solution to the discrepancy between the number of local group
dwarf galaxies and predictions from CDM models that do not include
outflows (\cite{kl99}).

\begin{figure}[t]
\vspace{10.1cm}
\includegraphics{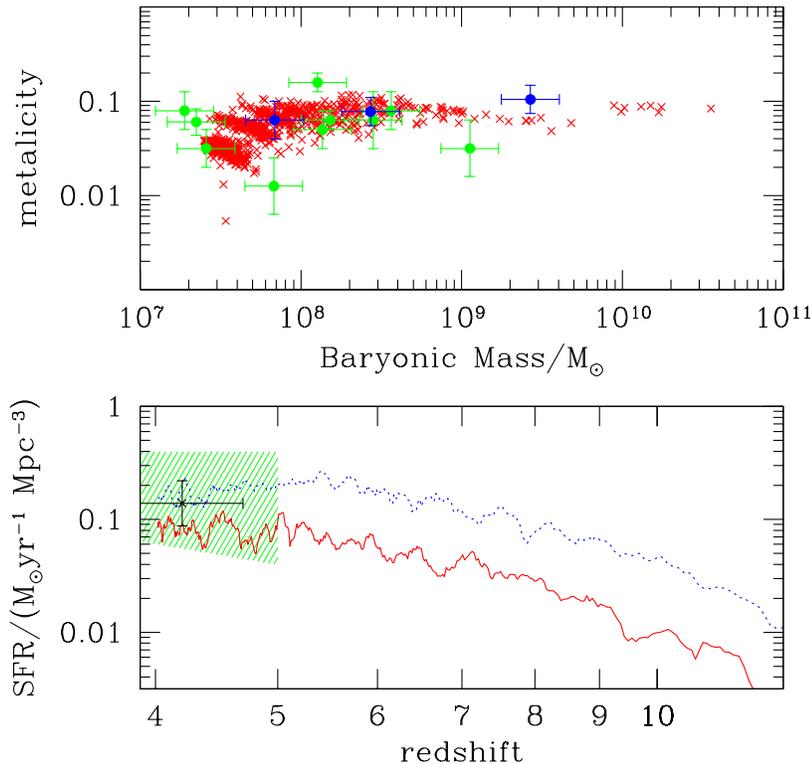}
\caption{{\bf (A)} Mean metallicity of gas particles within each
galaxy as a function of its total baryonic mass, at a fixed redshift
$z=4$ (crosses).  Here the solid points are local group dwarves
(\cite{ma98})
shifted by a constant factor of
$\Omega_b/\Omega_0.$ The green points represent observations of iron
abundances, while the blue points represent objects for which only the
oxygen abundance has been measured. {\bf (B)} The star formation rate
as a function of redshift.  The solid lines and dotted lines
correspond to the fiducial and no-outflows case respectively.  The
data point is taken from observations (\cite{st99}) and the green
region is a compilation of indirect constraints (\cite{ho01}).}
\label{fig:fig2}
\end{figure}

Above $\gtrsim 5\times10^9M_\odot$, the number-densities in the two
models become similar, which was not observed 
in our previous work (\cite{st01}). 
Thus this
feature may be related to the key difference between these
simulations: the inclusion of gas enrichment by supernovae.  To study
this further, we plot the mean metallicity of the gas particles
identified in galaxies as a function of galaxy mass in Figure 
\ref{fig:fig2}A. Here
we see that above $\sim 10^9 M_\odot$, the metallicity is roughly
constant at $0.1 \;Z_\odot$.  While our models do not include the
quiescent mode of star formation that is important in such large
objects, this value is suggestive of the $\sim 0.1 \;Z_\odot$
pre-enrichment necessary to explain the lack of low-metallicity G
dwarf stars in the solar neighborhood (\cite{os75}), as well as in other
large nearby galaxies (\cite{to99}), a result that is consistent with
semi-analytic studies (\cite{sc01}).

Using a simple estimate for the ``cooling time'' (see \cite{sc01})  we can 
get a rough idea of the delay between gas
collapse and cooling into a disk for the large objects in these
simulations, which is a strong function of metallicity. Referring to 
Figure
1, we see that above $2 \times 10^{10} M_\odot \Omega_0/\Omega_b$ the
enrichment run appears to contain more massive galaxies than the
non-enrichment run, although there are only two halos
in the simulation that are above this threshold and we would prefer better
statistics before making definitive conclusions. For the sake of
reference we call this mass threshold the `crossover' mass.
We note that there was no evidence for a crossover phenomenon in
our previous simulation work (Scannapieco \etal 2001).

For
galaxies with total masses roughly corresponding to the crossover value
the cooling time is $\sim 3.0 \times 10^8$ years for $0.05 \;Z_\odot$
metallicity gas and $1.5 \times 10^8$ years for gas at $0.1 \;Z_\odot$ 
metallicity.  Comparing these values with the age of the universe in this
cosmology ($11.5 \times 10^8$ years at $z=5$ and $14.8 \times 10^8$ years
at $z=4$) shows that approximately $3.3 \times 10^8$ years passes between
$z=4$ and $z=5$, which is very slightly larger than the cooling time for
the $0.05 \;Z_\odot$ gas, and about twice the cooling time of the $0.1
\;Z_\odot$ gas. Both of these cooling times are significantly shorter than
the Hubble time at $z=4$ which implies that the increased cooling rate can
only have an effect on halo gas that is close to the galaxy. Given the
small mass fraction of this gas it seems somewhat unlikely that it would 
contribute to such a noticeable mass increase.
An alternative explanation may be that massive halos act as a sink for
outflow material from surrounding smaller systems. However, if this were
the case we would expect to have seen the effect of this in Scannapieco et
al. (2001). Ultimately, an exact determination of the nature of the  
apparent increase   
in mass requires a larger simulation box to improve statistics.

Our outflow model also has important implications for galaxies with
baryonic masses less than $10^9M_\odot$, whose metallicities range
from $0.02 \;Z_\odot$ to $0.1 \;Z_\odot$ in our simulations.  This
distribution can be compared with metallicities observed in local
group dwarf galaxies, which are thought to form at high redshift.  To
facilitate this comparison, we overplot in Figure \ref{fig:fig2}A the masses 
and
metallicities of these objects as compiled in a review of the 
local group galaxies (\cite{ma98}), using the
iron abundances when available, and assuming a ``shifted'' solar
abundance of $10^{9.4}$ in cases in which only the oxygen abundance
was measured.  Note that this standard shift is also taken in the
observational literature (\cite{ma98}) because these objects are rich in
$\alpha$ elements such as oxygen, which are abundant in the type-II
SNe such as those driving our outflows.  In each case we assume a
standard error in the mass determination of $50\%$, due to errors in
determining the velocity dispersion and distance to these objects, and
apply an overall mass shift of $\Omega_b/\Omega_0.$ Clearly our model
does a good job of reproducing the metallicity-luminosity relation
seen in dwarf galaxies, in which metallicity increases as a function
of luminosity, but with a scatter of almost an order of magnitude.
 
 Compared to the enrichment values presented in Gnedin \& Ostriker (1997)
our values are lower, by an approximate factor of 3-4, in the highest
density regions. Most of this difference is probably attributable to the
difference between the star formation rates, as the inclusion of outflows
systematically lowers our SFR (although we lack a photoionization
background). We also have significantly less scatter in the observed metal
fractions, which is a function of differences between the numerical
methods employed. Gnedin \& Ostriker adapt the method presented in Cen \&
Ostriker (1992) which involves converting mass in cells obeying
overdensity, cooling and convergence criteria to collisionless stellar
particles. This can lead to a widely varying enrichment history depending
upon the details of the density of the gas in the protogalaxy. In our
model we ignore the inner structure of our halos and instead treat them as
uniform systems that undergo global starbursts when sufficient gas has
been accumulated, as is done in semi-analytic models. In passing, we
mention that our results for protogalaxy metallicity show good agreement
with averaged enrichment values presented in Cen \& Ostriker (1999),
provided that a log-linear extrapolation to z=4 is reasonable.

Despite the increased formation of large objects in the outflows
simulation, their comparative rarity at these redshifts causes the total
amount of cooled gas to be dominated by the suppression effect. In Figure
\ref{fig:fig2}B we plot the overall star formation rate (SFR) as a
function of redshift in each of these simulations. The plot shows that the
inclusion of outflows results in a decrease in the SFR by approximately a
factor of two. We also compare the calculated SFRs with direct
observations of star formation (\cite{st99}) and a broad range of indirect
constraints (\cite{ho01}).  While both models are consistent
with the indirect constraints, the direct observations are slightly
over-predicted by the no-outflows run and slightly under-predicted in the
fiducial run.  As our starburst model does not include the quiescent mode
of star formation, our results should be considered lower limits on the
global SFR.  Thus, although the data seems to favor the fiducial run, no
definitive conclusions can be drawn. 

It is worth noting that the star formation rate only becomes significant
at redshifts $\lesssim 10$.  This redshift is determined by the
minimum mass scale in our simulations, chosen to be the
minimum mass that can cool effectively without molecular hydrogen,
which we assume to be completely dissociated.  

We have not included a photoionization background in our simulation
primarily because we wanted to observe the effect of the outflow model
without having to disentangle it from other physical phenomena.
Nonetheless, it is important to assess what effect the inclusion of
photoionization field might have. Photoionization is known to reduce the
number density of dwarf systems, and is considered to be the primary   
solution to the `substructure problem' (\cite{kl99}; \cite{mo99}). Most
simulations of photoionization have concentrated on either large scales
with a comparatively low mass resolution (eg \cite{ka96})  or very
small scales
at high mass resolution to examine phenomenology of individual halos (eg
\cite{ns97}). On the other hand semi-analytic studies are
able to
provide both good statistics and high mass resolution (eg \cite{so02}), at
the cost of lost geometric information and highly phenomenological
physics.

Our minimum halo mass resolution is slightly under $1.5\times10^8
M_\odot$. At this mass scale the $z=0$ V-band luminosity function for the
semi-analytic model of Sommerville (2002) is about a factor of three times
lower when a photoionization background is included. This result is
surprisingly insensitive to the actual photoionization epoch, and large     
differences between different photoionizing models are only found at mass   
scales that we do not resolve (around $10^7 M_\odot$, i.e. close to the 
mass of the Draco dwarf spheroidal). Thus, assuming that not all dwarf
systems are late forming, we can expect our simulation to over-predict the
number of dwarf halos by at most a factor of three. The natural effect of
a reduced number density of gaseous halos is a reduced global SFR and
reduced enrichment volume factor, although the metallicity values for 
individual galaxies will be similar. However, in all likelihood, some of
the protogalaxies that are destroyed by outflows would be suppressed, or
possibly erased, by a photoionization background.
 Thus an accurate assessment of the effect of a
photoionization background on the outflow model is quite difficult.
Also if the epoch
of reionization is optimistically late ($z\sim 7$), then we can expect a
less significant
impact on our model, as a large amount of the evolution will occur without
the presence of a photoionizing background. In this case the SFR at $z=7$
will undoubtedly plateau, or turn off somewhat, and the enrichment
fractions will be reduced. Ultimately, determining the precise impact of a
photoionization background requires running a self consistent simulation
that includes it.

\section{Outflows and the Intergalactic Medium}

The widespread generation of outflows has an impact over regions much
larger than the dense peaks identified as galaxies. Indeed the
strongest inference of an epoch of galaxy outflows is drawn from
observations of the IGM itself. 

\begin{figure}[t]
\vspace{10.3cm}
\includegraphics{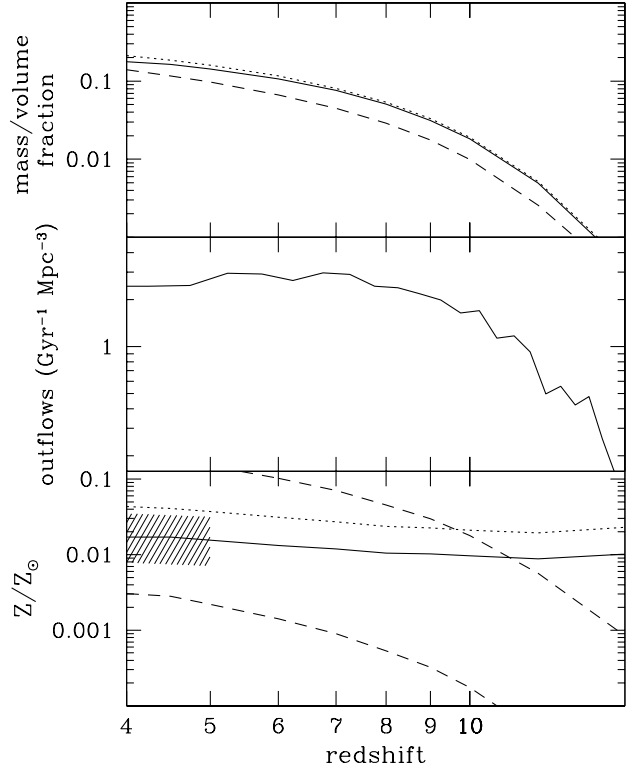}
\caption{{\bf (A)} Fraction of gas particles enriched by outflows
as a function of redshift.  Here the solid lines represent the mass
fraction of particles that are {\em not} labeled as belonging to
galaxies, while the dotted lines include all enriched gas particles.
Finally, the dashed line gives that volume fraction enriched to
$0.0001\;Z_\odot$ or greater.  {\bf (B)}
Outflows rate per unit volume as a function of redshift.  {\bf (C)}
Mean metallicity of enriched particles as a function of redshift.
Solid and dotted lines are as in A, while the dashed line is an
average over all gas particles in the simulation, not tagged as
galaxies.  The green region shows the range of metallicities for the
gas within galaxies as calculated in earlier analytic
models (\cite{pi99}).}
\label{fig:fig3}
\end{figure}

There are two methods by which this impact can be quantified.  The
first of these is by examining the overall mass of gas enriched by
stellar material, which we plot as a function of redshift in Figure
\ref{fig:fig3}A.  While outflows vigorously enrich the densest
regions, the IGM transition associated with outflows is incomplete,
and only $\sim 20\%$ of the particles are enriched by redshift 4.  In
this plot we also see that the majority of enrichment takes place at
redshifts below 10, when atomic transitions become an effective
coolant.

The onset of this enrichment roughly corresponds with the sharp rise
in the number of outflows at $z>10$, as shown in Figure
\ref{fig:fig3}B.  This plot, which displays the outflow rate per
unit Mpc$^3$ per Gyr, also hints at the end of the epoch of galaxy
outflows as the curve begins to turn over at lower redshift values.
The details of this turn-over are quite uncertain, however, and model
dependent.  The determining factor in this case is the maximum size of
the gravitational potential from which SNe can successfully cooperate
to drive a large-scale outflow, a quantity that is crudely modeled by
our overall upper limit of 1000 particles.  In reality this turn-over is
dependent on such uncertain quantities as the distribution of
high-redshift OB associations, the star formation efficiency, and the
stellar initial mass function in high-redshift galaxies (\cite{mo01}).

In Figure \ref{fig:fig3}C we examine the overall level of
enrichment within the affected regions, and also plot the bounds from
analytic models (\cite{pi99}) for reference.  The slight 
decrease in these
values seen at large redshifts is likely to be due to low-number
statistics, as only $\sim 15$ objects are formed in this simulation by
$z=15$.  Despite the fact that outflows are able to enrich the largest
objects to metallicities of $\sim 0.1\; Z_\odot$, their mean impact on
polluted regions of the IGM is only on the order of $0.02\; Z_\odot$, and
the mean value when averaged over the simulation as a whole is $0.003\; 
Z_\odot$.  Thus our model not only results in a patchy distribution of
enriched regions, but also a large variance within these regions.
This is consistent with observations of \CIV at $z \approx 3$ in QSO
absorption systems, which can be reproduced by a model in which the
overall mean metallicity of the IGM is $0.003\; Z_\odot$ with a large scatter 
of
roughly an order of magnitude (\cite{he97}).

\begin{figure}[t]
\vspace{11.2cm}
\includegraphics{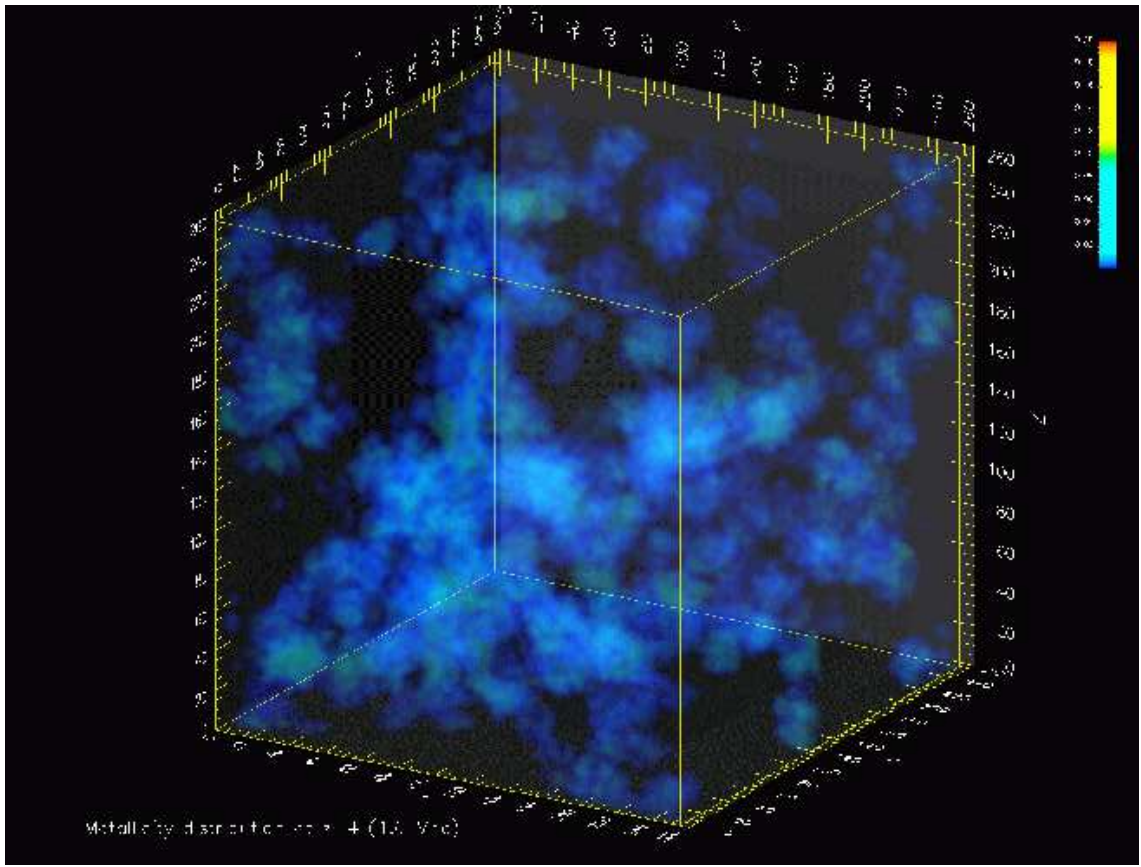}
\caption{Volume-rendered image of metal enrichment at
$z=4$. (Full resolution version available from 
coho.mcmaster.ca/$\, \tilde{}$ thacker/fig4.eps.gz.)}
\label{fig:fig4}
\end{figure}

To get a better handle on this inhomogeneity, we turn our attention to
the second method of quantifying this IGM transition and examine the
volume impacted by outflows.  Figure \ref{fig:fig4} is a volume-rendered 
image of
the smoothed distribution of metals in our simulation volume.  Here we
see that the spatial distribution of material from outflows is highly
biased, illustrating the necessity of studying this problem through
simulations or other techniques that capture this inhomogeneity 
(\cite{sc01}).
  While much of the simulation remains metal free, large
clouds of $\sim 0.02 \;Z_\odot$ material roughly outline the underlying
filaments of dark matter, punctuated by smaller regions of higher
metallicity.

In Figure 3A we plot the redshift evolution of the volume     
enriched to
0.001$Z_\odot$ or greater. The volume fraction enriched follows the trend
of the mass enriched but is reduced because of the large overdensities at
which galaxies form.  To quantify this further, in Figure 5 we plot the   
volume filling factor of material with metallicity above a given threshold
value, plotted as a function of the threshold value at $z=4$.  At this   
redshift the maximum enrichment volume is slightly over 20\%, consistent
with new semi-analytic calculations (\cite{ef02}),
and apparently at odds with the claimed (\cite{mad01})
100\% value for $z=10$.

While Madau, Ferrara, \& Rees (2001) only considered outflows from
objects at a fixed mass scale of of $10^8 M_\odot/h$ and then
multiplied by a Press-Schechter estimate of their number density
to compute to overall filling factor, the overall energetics of their
wind models are quite similar to ours.  Thus it is likely that the   
primary source of this inconsistency is related instead to their
adoption of an sCDM cosmology with parameters $\Omega_m=1$, $h=0.5$,
$\sigma_8=0.63$, $n=1$, $\Omega_b h^2=0.019$. Indeed, using the
standard Press-Schechter approach to calculate the comoving number
density of halos of masses greater than $10^8 M_\odot/h$ at $z=9$ we
find that the sCDM model of Madau, Ferrara \& Rees (2001) contains 5.5
times as many objects per Mpc h${}^{-1}$ as our $\Lambda$CDM model.
This enhancement is primarily due to the association of a smaller
top-hat filtering scale to sCDM objects as compared with their
$\Omega_M = 0.35$ counterparts, and is sufficient to explain the factor
of 5 difference that we observe between our results. We therefore
conclude that the difference between filling factors can be attributed
almost entirely to the reduced number counts of dwarf galaxies in the
observationally motivated $\Lambda$CDM cosmology compared to the out
of favor sCDM model. 

\begin{figure}[t]
\vspace{8.6cm}
\includegraphics{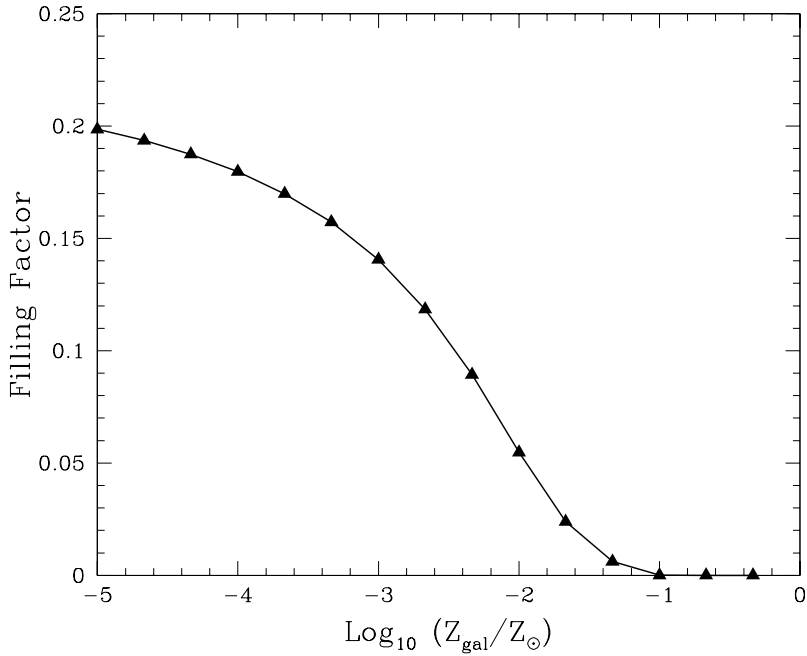}
\caption{Volume filling factor of enriched material
with metallicity values above a given threshold, plotted
as a function of threshold at $z=4$.}
\label{fig:fig5}
\end{figure}

Nonetheless, Figure 5 clearly shows that enrichment is far from
uniform.  Metallicities greater than $0.01\;Z_\odot$ are confined to
less than 5.5\% of the volume, while enrichments in excess of $0.1
\;Z_\odot$ are confined primarily to larger dwarf galaxies and their
halos, which occupy less than 0.1\% of the simulation volume.  This
correlation between the distribution of metals and starbursting
galaxies provides a natural explanation for a long standing dichotomy
seen in QSO absorption line studies.  For while it is clear that the
distribution of Ly$\alpha$ in the IGM as a whole is well modeled
without considering outflows (\cite{mc00}), studies of \CIV clouds
indicate that these enriched objects have been turbulently stirred by
events occurring on timescales on the order of $10^8$ years (\cite{rs01}).

\section{Conclusion}

In this work, we have studied the role of outflows and metal
enrichment in structure formation, comparing our results to a diverse
set of observations. We have not attempted to calculate the properties
of individual galaxies, but rather have used results from previous
studies as input into our model.  While based on the approach described
in Scannapieco, Thacker, \& Davis (2001), this study is a significant
advance over our previous work as it includes a more realistic
model of metal enrichment as well as a more efficient SPH algorithm
that allows us to integrate to directly observable redshifts.
This leads to a number of important conclusions, which merit
further investigation.

First, while the presence of outflows greatly reduces the number of
galaxies formed at lower mass scales as a result of baryonic stripping, this
mechanism is effective only for galaxies with total mass below
$5\times10^{9}$ M${}_\odot$.  At greater mass scales, the galaxy
number counts in the outflows case become similar or even exceed that
of the no-outflows case as a result of more rapid cooling caused by
metal enrichment.

Second, the inhomogeneous distribution of metals by outflows leads
naturally to a scatter in metallicity of lower mass objects that 
is consistent with observations of local group dwarf galaxies.
Above $10^{9}$ M${}_\odot$, however, the metallicity is roughly
constant at 0.1 Z${}_\odot$ in agreement with pre-enrichment estimates
necessary to explain the lack of low-metallicity G-dwarf stars in the
solar neighborhood.

Third, the global SFR for our model is in broad agreement with 
observations, albeit on the low side. This is attributed to the lack of 
quiescent star formation in our model.

Finally, our global enrichment value of 0.3\% Z${}_\odot$ is 
consistent with  observations of \CIV at z$\simeq$3 in QSO absorption   
systems  This metal enrichment is far from uniform and concurs with the 
interpretation that \CIV clouds are turbulently stirred on small scales
while the distribution of Ly$\alpha$ clouds remains comparatively 
unaffected.

Future work needs to address the role of H${}_2$ cooling in our
simulations as at the moment we make no attempt to calculate
production of H${}_2$. Progress is difficult on this front as while
calculation of rate equations is comparatively straightforward (and
has been conducted within in our simulation framework, see
\cite{fc00}) a fully consistent solution with local
photodisassociation is, at least at the moment, an extreme
computational challenge (eg \cite{rs99}). We also plan to include the
fiducial photoionization background associated with the epoch of
reionization. We will address both of these issues over the coming
months, although we do not anticipate a major revision of our current
results.

\acknowledgments
We thank Andrea Ferrara, Tom Broadhurst and Mordecai Mac Low
for helpful discussions, Greg Bryan for making available
unpublished results and the anonymous referee for suggesting a number of 
changes 
that significantly improved the content of the paper.
We also gratefully acknowledge the hospitality
of Carlos Frenk and the Physics Department at the University of Durham
where this paper was finished. E.S.\ has been funded in part by an NSF
MPS-DRF fellowship. R.J.T.\ acknowledges funding from the Canadian
Computational Cosmology Consortium and NSERC of Canada via the 
operating grant of Prof H. M. P. Couchman. This project was supported
by NSF KDI grant 9872979.

\end{document}